\documentclass[fleqn,10pt]{wlscirep}
\usepackage[utf8]{inputenc}
\usepackage[T1]{fontenc}
\usepackage{bm}

\usepackage{amssymb}
\usepackage{mathptmx}
\usepackage{amsmath}	
\usepackage{gensymb}
\usepackage{color}
\usepackage{enumitem}

\usepackage{epsfig}
\usepackage{color}
\usepackage{graphicx}
\usepackage{longtable}
\usepackage{hyperref}

\usepackage[labelsep=endash]{caption}

\newcommand{\msun}{M$_{\odot}$}
\newcommand{\lcdm}{$\Lambda$CDM}

\newcommand{\mstar}{{\rm M_{*}}}

\newcommand{\kms}{{\rm km.s^{-1}}}
\newcommand{\vmax}{{\rm V_{max}}}
 
\title{Baryonic solutions and challenges for cosmological models of dwarf galaxies}

\author[1,*]{Laura V. Sales}
\author[2]{Andrew Wetzel}
\author[3]{Azadeh Fattahi}
\affil[1]{Department of Physics and Astronomy, University of California, Riverside, CA 92507, USA}
\affil[2]{Department of Physics and Astronomy, University of California, Davis, CA 95616, USA}
\affil[3]{Institute for Computational Cosmology, Department of Physics, Durham Univ., South Road, Durham, DH1 3LE, UK}

\affil[*]{e-mail: lsales@ucr.edu}

\begin{abstract}
Galaxies and their dark-matter halos have posed several challenges to the Dark Energy plus Cold Dark Matter (\lcdm) cosmological model. These discrepancies between observations and theory intensify for the lowest-mass (`dwarf') galaxies. \lcdm\; predictions for the number, spatial distribution, and internal structure of low-mass dark-matter halos have historically been at odds with observed dwarf galaxies, but this is partially expected, because many predictions modeled only the dark-matter component. Any robust \lcdm\; prediction must include, hand-in-hand, a model for galaxy formation to understand how baryonic matter populates and affects dark-matter halos. In this article, we review the most notable challenges to \lcdm\; regarding dwarf galaxies, and we discuss how recent cosmological numerical simulations have pinpointed baryonic solutions to these challenges. We identify remaining tensions, including the diversity of the inner dark-matter content, planes of satellites, stellar morphologies, and star-formation quenching. Their resolution, or validation as actual problems to \lcdm, will likely require both refining galaxy formation models and improving numerical accuracy in simulations.
\end{abstract}

\begin{document}

\flushbottom
\maketitle

\thispagestyle{empty}

Baryonic matter constitutes only $\sim 17\%$ of the total mass budget in the Universe \cite{Planck2018} but it dominates what we call galaxies in observations. Therefore, modeling the effects of baryons is unavoidable to achieve a successful cosmological galaxy formation theory to compare against observations \cite{Rees1977, White1978}. The relevant physical processes in galaxies interact non-linearly with each other and also may back-react onto the (dominant) dark-matter component through gravity.
Thus, cosmological numerical simulations have emerged as powerful tools to follow the assembly of galaxies within dark-matter halos \cite{Somerville2015, Vogelsberger2020}.

In this Review, we focus on theoretical insights from cosmological baryonic simulations within $\Lambda$CDM on the formation of low-mass (`dwarf') galaxies, with stellar masses $M_* \lesssim 10^9$ \msun. Other theoretical approaches, such as analytical/semi-analytical methods \cite{Benson2010,DeLucia2019} and semi-empirical/forward-modeling techniques \cite{Conroy2006,Kravtsov2013,Moster2018,Somerville2018,Behroozi2019, Nadler2019} are also immensely valuable and complementary, though we refer the reader to the references above.
Furthermore, in this review we focus only on cold dark matter (CDM) as a viable dark-matter model. However, some tensions and challenges with observations might be mitigated, sometimes arguably more naturally, by changing the underlying nature of dark matter or modifying the law of gravity. We refer the reader to \cite{Bullock2017, Milgrom2002, Maeder2020} for a discussion of these approaches.

\section*{The physics of dwarf galaxy formation}

The formation of dark-matter structures in $\Lambda$CDM is a process relatively well understood, where halos form from the hierarchical growth of high-density fluctuations in an otherwise homogeneous early Universe. Halos assemble `hierarchically': low-mass halos collapse first and then merge to form more massive ones. Because CDM is assumed collisionless, only the effects of gravity are important to study the formation of dark-matter structures. Baryons, on the other hand, which initially were primordial gas, but then (in part) converted to stars and metals, decoupled early from the dark matter; modeling their evolution requires a complex network of physical processes, including hydrodyamics and the cooling and heating of gas, in addition to gravity. We refer to these as `baryonic processes'.

Several baryonic processes are essential to form realistic galaxies within \lcdm. An important aspect of their combined effects is a suppression of the efficiency of star formation, achieved from a combination of stellar feedback channels, including supernova explosions \cite{Dekel1986, Stinson2007,Sawala2010, Governato2010,Vogelsberger2014c, Gonzalez-Samaniego2014, Hopkins2018a} and radiation and winds from young stars \cite{WangDutton2015, Hopkins2020a}.
The extragalactic UV background, which drives cosmic reionization, suppresses gaseous accretion into galaxies and, on the extreme scales of ultra-faint dwarf galaxies ($M_* \lesssim 10^5$ \msun; see ref. \cite{Bullock2017}), cosmic reionization is thought to halt star formation entirely, making such present-day galaxies `fossils' of reionization \cite{Bovill2009, Bovill2011, Brown2014, Fitts2017, Revaz2018,Wheeler2019}. Although these processes all affect massive galaxies like the Milky Way (MW), dwarf galaxies, with their shallower dark matter potentials and fewer number of stars, are particularly susceptible to the physics of stellar feedback and reionization. Thus, dwarf galaxies are particularly sensitive laboratories for testing galaxy formation models.

Environmental effects also shape the dwarf galaxies that orbit inside a more massive host halo, which for MW-mass halos corresponds to distances $\lesssim 300 - 400$ kpc.
These `satellite' dwarf galaxies show differences in their properties compared with `isolated' (or `field' or `central') dwarf galaxies that are not embedded within a larger host halo. As they orbit, satellites experience significant tidal stripping from the host halo potential, leading to significant mass loss \cite{Gao2004, WetzelWhite2010}.
This stripping proceeds primarily outside-in, so it initially impacts the more extended dark-matter, only later affecting the more centrally concentrated, and more tightly bound, stars and gas in the galaxy \cite{Penarrubia2008, Choi2009, Libeskind2011, Brooks2013, Errani2015, Smith2016}. Simulations typically find that present-day satellites of MW-mass halos retain on average $20 - 40\%$ of their initial dark-matter mass and $\gtrsim 75\%$ of their stellar mass \cite{Sales2007a, Buck2019, Mazzarini2020}.
Eventually, the inner (luminous) region of a satellite can start to be stripped as well, which may help explain the kinematics observed for satellites of the MW \cite{Brooks2014, Sawala2016, Wetzel2016}.

After infall, the gas content of satellites also may get suppressed.
First, the host halo can prevent new accretion from the cosmic web, and eventually, ram-pressure via interaction with the host halo's gaseous corona can remove dense gas from a satellite \cite{Gunn1972, Abadi1999}, which in turn can turn off (`quench') star-formation. The modeling of gas content in satellite dwarf galaxies is necessary to produce realistic color gradients, quiescent fractions, and star-formation histories \cite{Font2008,Sales2015,Benitez-Llambay2015a,Wetzel2016,Buck2019, GarrisonKimmel2019b, Wright2019, Digby2019, Joshi2021}. Additional environmental effects such as tidal heating \cite{Gnedin1999}, tidal stirring \cite{Kazantzidis2011, Kazantzidis2017}, and biased formation in the higher-density environment \cite{Mistani2016,Artale2018,Joshi2021,Jackson2021} may help explain the different range of stellar sizes and morphologies in satellites compared to similar-mass central dwarf galaxies.

\section*{Tensions and problems of \lcdm: an updated view}

$\Lambda$CDM, a mature theoretical framework, has evolved through different phases and challenges. Our goal is to review historical so-called `problems' of$\Lambda$CDM on the scales of dwarf galaxies, describe how the additional computational modeling of baryonic physics at sufficiently high resolution has resolved or recast many of these `problems', and discuss ongoing challenges and sources of tension for models of $\Lambda$CDM that include the baryonic physics. Thus, we seek to recast these historical `problems' in a more productive and rigorous context.

In our evaluation, strictly speaking, a legitimate `problem' between theory and observations exists only if (1) a theoretical model that includes the relevant physics makes a firm prediction, and (2) a robust observational measurements disagrees with this prediction at a meaningful level (several sigma). In this sense, mere uncertainty, either in observations or theoretical predictions, does not a priori constitute a `problem'. Rather, uncertainty points towards interesting directions to pursue to test models more rigorously and assess whether a legitimate disagreement exists, given better observations, better theoretical understanding, or both.

The most famous example of a problem that has now been resolved is `missing satellites' \cite{Klypin1999, Moore1999}: that \textit{dark-matter-only} $\Lambda$CDM cosmological simulations of MW-mass halos predict many more satellites (dark-matter subhalos) than observed dwarf galaxies around the MW or Andromeda (M31).
In retrospect, several sources of uncertainty and incompleteness limited a robust comparison between theory and observation, including: (a) simulations not modeling the role of baryons and the formation of a MW-mass galaxy, (b) uncertainty in the relation between the dark-matter mass of a subhalo and its (observable) galaxy mass/luminosity, (c) limited numerical resolution, and (d) observational incompleteness in the number of satellites around the MW and M31. Indeed, two decades later, progress in both observations---with discoveries of dozens of new faint satellites \cite{Simon2019}---and improved theoretical models that directly predict observable properties of dwarf galaxies (like stellar mass/luminosity) has shown that \textit{there simply is no `missing satellites' problem}: current $\Lambda$CDM cosmological baryonic simulations at sufficiently high resolution are consistent (within reasonable theoretical and observational uncertainties) with the observed numbers of satellites around the MW and M31 \cite{Sawala2016, Wetzel2016, Simpson2018, Buck2019, Garrison-Kimmel2019, Munshi2021, Font2021, Engler2021b}, as we discuss below.

\begin{figure}[ht]
\centering
\includegraphics[width=\linewidth]{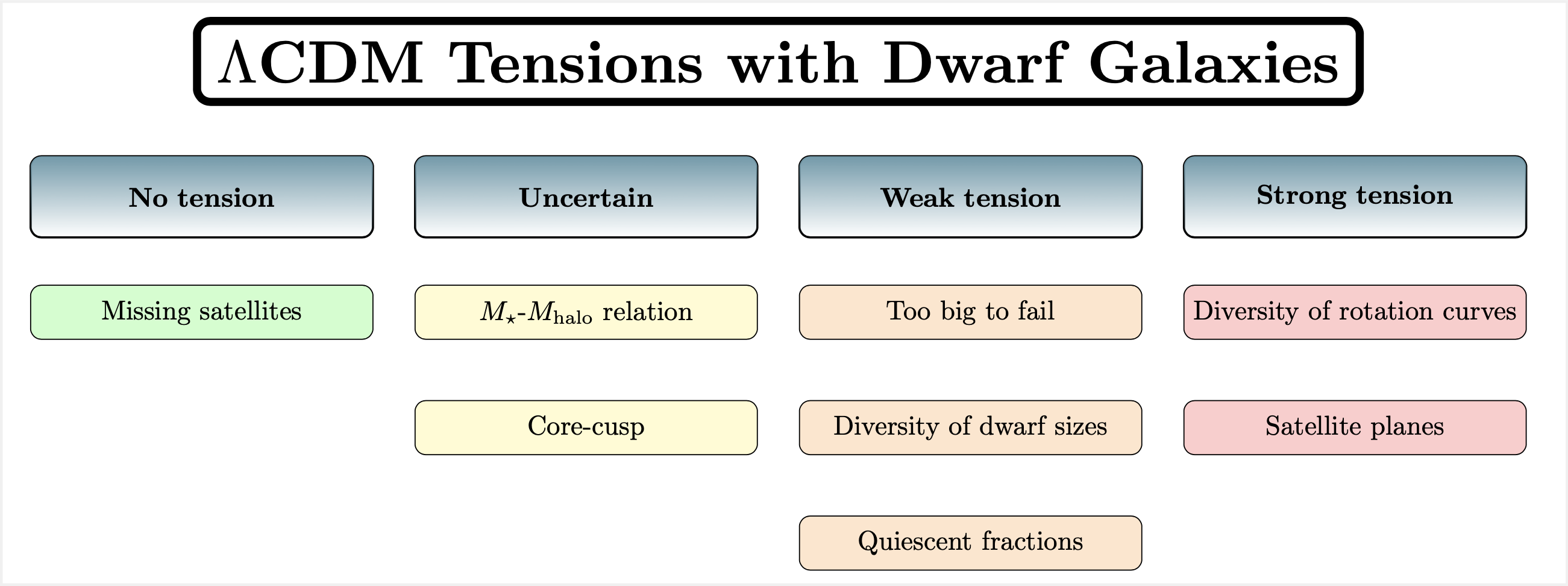}
\caption{{\bf Historical and current tensions between $\Lambda$CDM theory and observations of dwarf galaxies.}
 We classify these according to the level of tension/challenge they present to the cosmological $\Lambda$CDM scenario, \textit{after the critical effects of baryonic physics have been considered}. 
Left to right moves from `no tension' to `strong tension'.
The following sections discuss each of the topics in this chart. We discuss the $M_*$-$M_{\rm halo}$ relation and the Too-big-to-fail problem in sections with those respective names. We address the core-cusp problem and the diversity of rotation curves in the section: `Dark matter distribution within dwarf galaxies'; the diversity of sizes in the section: `Baryonic distribution within dwarf galaxies'; and satellite planes together with quiescent fractions grouped in the section: `Satellite dwarf galaxies'.
}
\label{fig:fig1}
\end{figure}

That said, several ongoing challenges persist and need to be addressed, and we propose to re-cast these according to the degree of `tension' between current theoretical predictions of $\Lambda$CDM that include baryons and robust observations of dwarf galaxies. In some cases, the baryonic solutions that address some of the traditional `problems', like missing satellites, might cause (or exacerbate) other tensions. In Figure~\ref{fig:fig1}, we list both historical and new tensions for $\Lambda$CDM, categorizing them by our evaluation of their current severity. We discuss them individually below.

\section*{Relation between stellar mass and dark-matter halo mass of dwarf galaxies}
\label{sec:SMHM}

The $\Lambda$CDM model makes clear predictions for the mass function of dark-matter halos \cite{Gao2004,Springel2008}. Predictions for the counts of faint dwarf galaxies then follow from knowing the relation between stellar mass and halo mass. However, dark-matter halo masses are challenging to measure observationally. Instead, the luminosity and stellar mass functions of galaxies have been of paramount importance for validating cosmological models. A challenge is that the counts of ultra-faint galaxies, down to $M_* \sim 100 - 1000 \; \rm M_\odot$, remain mostly unconstrained, even within the MW halo \cite{Simon2019}. Therefore, currently it is challenging to evaluate if theoretical predictions agree with observations.

Alternately, on just the theoretical side, one can compare the predictions of different simulations regarding the relation between galaxy stellar mass and dark-matter halo mass in the ultra-faint regime. Indeed, as discussed below, a careful look into state-of-the-art numerical simulations that predict the correct number of MW-like galaxies and classical dwarf galaxies suggests that their expected ultra-faint populations may differ, signaling an important theoretical uncertainty that persists. We thus emphasize that our discussion of this relation between stellar mass and dark-matter halo mass is different from the others in this review, because our comparison is only between different simulations, not (yet) between simulations and observations.

\begin{figure}[ht]
\centering
\hspace{-0.5cm}
\includegraphics[width=1.0\linewidth]{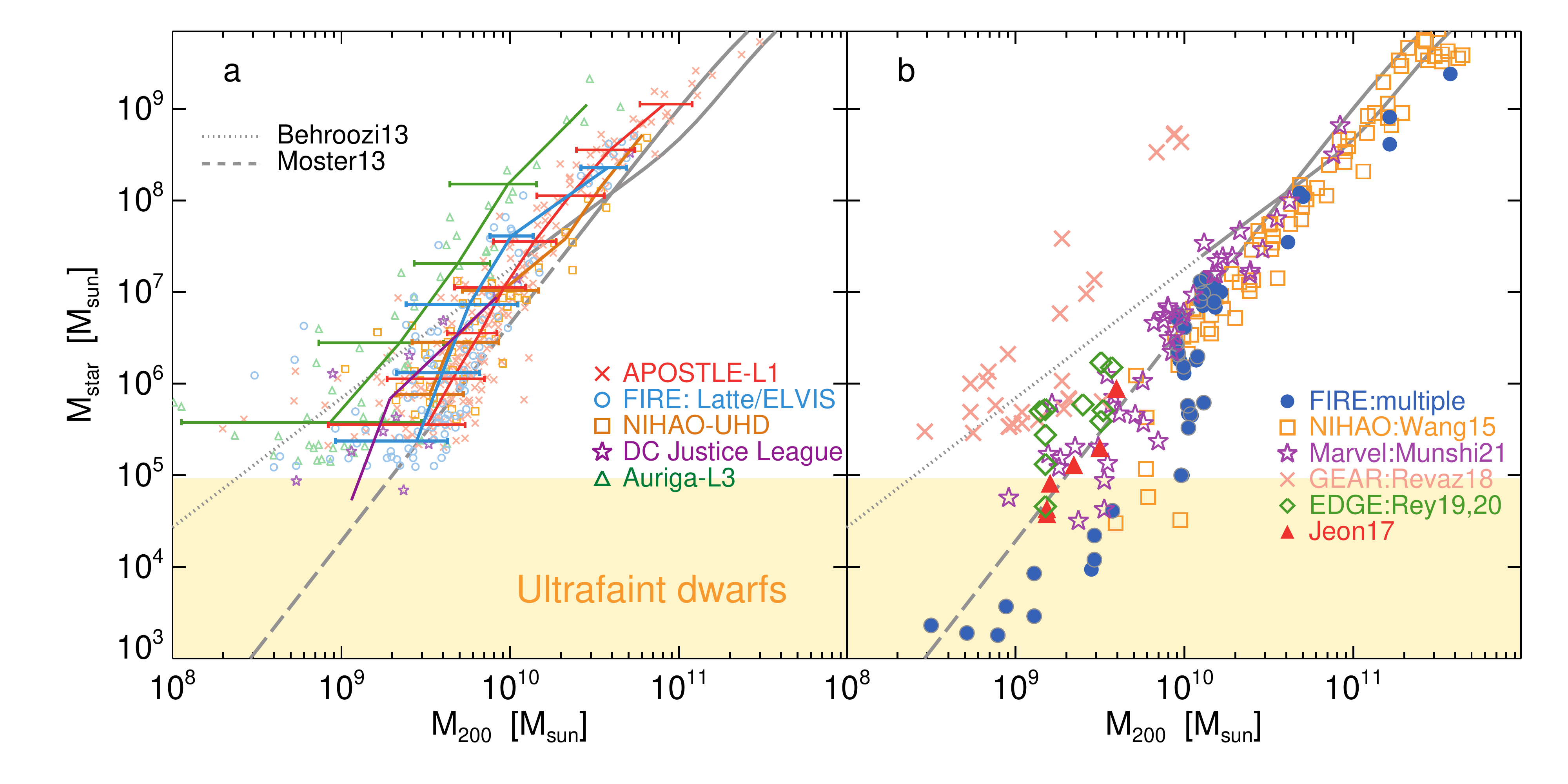}
\caption{
\textbf{Relation between galaxy stellar mass and dark-matter halo mass.} Central/field (non-satellite) dwarf galaxies in a sample of state-of-the-art cosmological simulations of galaxy formation are presented, as well as abundance matching models of ref. \cite{Moster2013} and \cite{Behroozi2013}.
{\it Panel $(a)$}: Central/field dwarf galaxies, beyond a MW-mass host halo, in simulations of MW-like or Local Group-like environments from various models; APOSTLE, L1 resolution \cite{Sawala2016,Fattahi2016} from the EAGLE project \cite{Schaye2015} shown as red crosses; the Latte and ELVIS suites \cite{Wetzel2016, Garrison-Kimmel2019} of FIRE-2 simulations \cite{Hopkins2018} shown as blue circles; Auriga \cite{Grand2017} shown as green triangles; NIHAO-UHD \cite{Buck2019} as orange squares; and DC Justice League \cite{Munshi2021} as purple stars. Lines with corresponding colors show the median halo mass at fixed stellar mass and the [10$^{\rm th}$-90$^{\rm th}$] percentile.
We include simulated dwarfs with stellar masses above $\sim 20$ times the initial gas mass resolution corresponding to each simulation.
{\it Panel (b)}:
Simulations that zoom-in on individual dwarf-mass halos; blue circles show FIRE-2 \cite{Hopkins2018, Wheeler2015, Fitts2017, Wheeler2019}; orange squares present ref. \cite{WangDutton2015} from NIHAO; open green diamonds correspond to refs. \cite{Rey2019,Rey2020} from EDGE project; light red crosses show ref.  \cite{Revaz2018} results using GEAR code, red triangles those from ref. \cite{Jeon2017} and the Marvel simulations \cite{Munshi2021} in purple stars. Note that the Marvel Suite simulates a zoom-in volume with several isolated dwarf halos and therefore falls in-between the definitions for the left and right panels. Solid lines show abundance matching relations in the regime where they are constrained and switch to dotted \cite{Behroozi2013} and dashed \cite{Moster2013} where they are extrapolated. The yellow shaded regions in both panels indicate the ultrafaint dwarf regime.
}
\label{fig:smhm}
\end{figure}

Fig.~\ref{fig:smhm} shows the relation between stellar mass and dark-matter halo mass, where we collect the present-day relation predicted from a sample of state-of-the-art cosmological simulations. Halo mass corresponds to the spherical radius within which the average density is $200$ times the critical density, the so-called virial radius. Where a different definition of halo mass was presented in the published work, we convert those values using average mass-concentration relation from ref. \cite{Ludlow2016}. On the left panel, we include zoom-in simulations of MW-like or Local Group-like environments from various works: APOSTLE \cite{Sawala2016,Fattahi2016} from the EAGLE project \cite{Schaye2015}, Latte and ELVIS suites \cite{Wetzel2016, Garrison-Kimmel2019} from the FIRE-2 project \cite{Hopkins2018}, Auriga \cite{Grand2017}, NIHAO-UHD \cite{Buck2019}, DC Justic League \cite{Applebaum2021}; or zooms of relatively large regions, like the Marvel Suite \cite{Munshi2021}. In all cases, we show only central (field) galaxies (not satellites), which are located beyond a MW-mass halo within the zoom-in region and therefore have not been stripped of mass like satellites have.

The numerical resolution of these simulations varies between a gas particle mass $\sim 10^3$ \msun\; for the highest resolution case (Marvel Suite), $\sim 5\times10^3$\msun\, for Auriga-L3 and FIRE-2, to $\sim 10^4$ for APOSTLE and NIHAO-UHD. The physics modeled and its particular implementation also vary from code to code, often with differences in predictions far more impacted by these physics choices than by numerical resolution. A detailed and fair account of the physics included in each simulation is beyond the scope of this review. But each simulation included in Fig.~\ref{fig:smhm} is a good example of the current state of affairs in galaxy formation modeling with demonstrated successes in the prediction of MW-like galaxies with realistic sizes, morphologies, kinematics, metallicities, star-formation rates, among other properties.

There is substantial overlap on the space spanned by different simulations, which is encouraging given the different codes and hydrodynamical solvers involved. In general, models approximately follow the extrapolations (dotted/dashed lines) from abundance matching relations \cite{Behroozi2013, Moster2013} calculated from more massive galaxies. However, in detail, the slope and the scatter for the stellar mass - halo mass relation may differ for each simulation. For instance, for a halo mass with $M_{\rm 200c} \sim 3 \times 10^{10}$ \msun, simulations predict a dwarf galaxy within a stellar mass range spanning 1 dex, $M_*=10^8$-$10^9$ \msun\; despite the scatter intrinsic to each model being quite small for that halo mass. Conversely, for a dwarf galaxy with $M_*=[0.6,1.2] \times 10^6$ \msun, the median halo masses predicted may differ by a factor $\sim 4$ between different models.
We caution that a tight relation between halo mass and stellar mass with small scatter, used for abundance matching of more massive galaxies, might not hold true for dwarf galaxies, where the scatter is expected to be larger \cite{Sawala2016,Garrison-Kimmel2017a, Munshi2021}. However, this exercise highlights the level of variance expected in the stellar content at fixed dark-matter halo mass (and vice versa) between the different models.

Cosmological simulations can achieve higher resolution by zooming in on regions of individual dwarf galaxies instead of MW-like or LG-like hosts, which allows them to model the ultra-faint edge of galaxy formation. The right panel of Fig.~\ref{fig:smhm} includes zoom-in simulations of individual dwarf galaxies from different codes: refs \cite{Hopkins2018,Wheeler2015, Wheeler2019,Fitts2017} from FIRE-2, ref. \cite{WangDutton2015} from NIHAO, refs. \cite{Rey2019,Rey2020} from the EDGE project, ref. \cite{Revaz2018} using the GEAR code, and ref. \cite{Jeon2017} using a modified version of Gadget-2. Despite the higher resolution, the differences between codes intensify, with the predicted stellar mass differing by more than $\sim 2$ orders of magnitude for halo masses $M_{\rm 200c} \sim 10^9$ \msun or a factor $\sim 10$ in halo mass for $M_*\sim 10^6$ \msun. While the small number of simulations and different accretion histories may help explain some of the differences, Fig.~\ref{fig:smhm} confirms that the prediction for the relation between stellar and halo mass in the ultra-faint regime strongly depends on the simulation model.
Therefore, ultra-faint galaxies persist as one of the most sensitive laboratories for any model of galaxy formation.

Beyond central (field) galaxies, the stellar mass function of satellites also informs the stellar-halo mass relation, because the subhalo mass function (of halos within a more massive host halo) is a clear prediction of $\Lambda$CDM \cite{Springel2008,Bullock2017}. All simulations in the left panel of Fig.~\ref{fig:smhm} predict realistic luminosity/stellar-mass functions for satellite dwarf galaxies, at least at $M_{*} \gtrsim  10^5$ \msun, as compared with observations (not shown here, we refer the reader to the original papers for details). The low efficiency of galaxy formation discussed above plays a crucial role in reproducing realistic number of dwarf galaxies from the steeply rising number of low-mass dark-matter halos and subhalos predicted in $\Lambda$CDM \cite{Jenkins2001,Moore1999,Klypin1999, Springel2008, Giocoli2008, Yang2011}. However, the uncertainty in the relation between stellar mass and halo mass implies a substantial uncertainty in the predicted counts of ultra-faint dwarfs galaxies within MW-mass analogs \cite{Garrison-Kimmel2017a, Munshi2021}.

A related aspect of models in the ultra-faint regime is the halo occupation fraction: the fraction of halos at a given mass that host a galaxy (versus remain dark).
The heating of gas from the ultraviolet background during the epoch of reionization is thought to prevent the formation of galaxies below a certain halo mass \cite{Navarro1997_photoheating,Gnedin2000,Bullock2000,Somerville2002,Okamoto2008} while keeping the star-formation efficiency in ultra-faint galaxies low \cite{Benson2002, Hoeft2006, Okamoto2009,Fitts2017}. However, the details of reionization, including speed (fast/slow), time (early/late), and mode (homogeneous/patchy), combined with the particular assembly history of low-mass halos near the threshold of galaxy formation, create scatter in this transition from ultra-faint galaxies to completely dark halos \cite{Fitts2017,Benitez-Llambay2020,Rey2019,Rey2020}. Interestingly, although some halos might never have formed stars, they still might host gas in thermodynamic equilibrium with the cosmic ultraviolet background and therefore be detectable through atomic-gas surveys \cite{Benitez-Llambay2016}.

Current estimates for the maximum circular velocity below which halos remain dark are $V_{\rm circ} \leq 20$ km $\rm s^{-1}$. However, given the strong additional tidal stripping from the MW galaxy \cite{Donghia2010, Zolotov2012, Brooks2013, Garrison-Kimmel2017}, some works suggest that there are not enough subhalos above that velocity scale to host the observed population of ultra-faint galaxies around the MW \cite{Graus2019,Kelley2019,Nadler2019}.
This implies a need for lower-mass halos to form ultra-faint galaxies.
In other words, modeling the additional tidal effects of the MW baryonic disk strongly strips (and can effectively destroy) dwarf galaxies with small pericenters, provoking a possible paradigm shift from the previous `missing satellites' problem to an opposite tension of `not enough satellites'.
However, in our evaluation this is not yet a robust tension for $\Lambda$CDM, because these results require confirmation from higher-resolution simulations that are less affected by artificial numerical disruption \cite{vandenBosch2018a,vandenBosch2018b,Li2019,Bose2020,Errani2021}.
This controversy shows that our understanding of the early Universe and formation of ultra-faint galaxies remains under active development.
We therefore consider our understanding of the relation between stellar mass and dark-matter halo mass for dwarf galaxies as `uncertain', as we indicate in Fig.~\ref{fig:fig1}.

\section*{Dark matter distribution within dwarf galaxies} 
\label{sec:DMdistribution}

Early CDM-only simulations revealed dark-matter halos to be  `cuspy', with densities diverging as $\rho \propto r^{-1}$ in the inner regions \cite{Dubinski1991,Crone1994,Navarro1995c}. Once properly scaled, the density distribution of a halo of any mass can be parametrized by a single `NFW' profile with one free parameter \cite{nfw1996,nfw1997}. Although improved numerical resolution suggested later that Einasto profiles with two free-parameters \cite{Einasto1965} and an inner slope that asymptotically approaches $r^{-0.75}$ were a better description overall \cite{Navarro2004,Navarro2010}, cuspy NFW profiles are good enough representations of the halo regions accessible to galaxy observations \cite{Ludlow2013}.

This prediction is, however, often in tension with the slowly rising rotation curves observed in some dwarf galaxies, which suggest that their inner densities are more consistent with a constant-density `core' \cite{deBlok2008,Oh2011}. This conflict became known as the `core-cusp' problem \cite{Flores1994, Moore1994}, which commonly has been identified in gas-rich dwarf galaxies with luminosities $L \geq 10^7\; \rm L_\odot$. Cores also are inferred in some gas-free lower-mass satellite dwarf galaxies in the Local Group, based on the velocity dispersion of the stars \cite{Battaglia2008,Amorisco2012, Walker2011}, although the results remain controversial \cite{Strigari2017, Genina2018, Harvey2018}.
In practice, because measuring the exact shape of the mass profile in the inner region of the rotation/dispersion curve is challenging, it is more robust to phrase this as an `inner mass deficit' tension \cite{Weinberg2015,Oman2015,Oman2019}: CDM predicts more dark matter in the inner regions of dwarfs than inferred from observations.

\begin{figure}[ht]
\centering
\hspace{-1cm}
\includegraphics[width=0.95\linewidth]{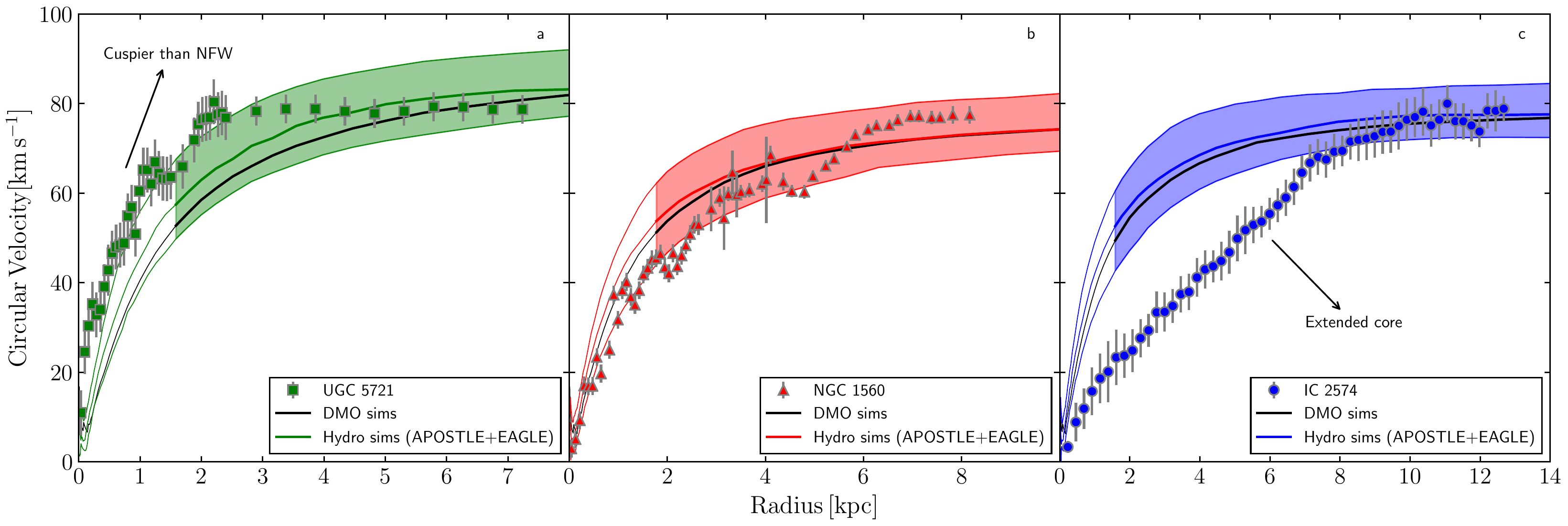}
\caption{{\bf Diversity of rotation curves, a persistent challenge to \lcdm.} Observed rotation curves of dwarf galaxies show a wide range of shapes in the inner regions. Here exemplified, data from three observed dwarfs (symbols with error bars) with similar outer rotation velocity $V \sim 80$ km $\rm s^{-1}$, but quite distinct inner behavior: from more steeply raising than NFW (a, UGC 5721), to well described by an NFW profile (b, NGC 1560), to showing a very extended core (c, IC 2574). Error bars account for statistical and systematic errors. Most baryonic simulations have been unable to recreate consistently the different velocity curve shapes in the inner regions without resorting to very strong observational biases. Thick solid color lines show the expectation (medians) from halos in the maximum circular velocity range $\sim 80$-$100$ km/s in the APOSTLE and EAGLE baryonic simulations with thin lines plus shading indicating $10^{\rm th}$-$90^{\rm th}$ percentiles (shading starts after the convergence radius, the minimum distance where results are presumed reliable). For comparison, black solid line shows a similar exercise using the dark matter only version. Although different codes have reported successes in forming cores in the inner regions (see text for details), reproducing cores and cusps has remained a challenge for modern galaxy formation simulations. {\it Figure adapted from ref. \cite{Oman2015}.}}
\label{fig:diversity}
\end{figure}

However, these are predictions from dark-matter-only simulations, and baryons can alter them. On the scale of dwarf galaxies, simulations show that stellar feedback can drive strong fluctuations to the gravitational potential by temporally driving gas out of the galaxy. Such potential fluctuations heat the orbits of dark-matter particles and effectively lowers the density of dark matter on the scales of the galaxy \cite{Navarro1996a, Gelato1999, Read2005, Pontzen2012}.

This scenario has a few key requirements. The potential fluctuations need to be non-adiabatic \cite{Navarro1996a}, on timescales shorter than the dynamical/orbital time, to heat the orbits of dark-matter particles and move them to more extended (larger apocenter) orbits, flattening the inner cusp to a core \cite{Pontzen2012}. Multiple `blow-out' episodes are more effective than a single one \cite{Read2005,Mashchenko2008,Governato2010,Pontzen2012,Teyssier2013}, which suggests that galaxies in which star-formation proceeds in several consecutive burst likely will have larger cores. However, burstiness alone is not a sufficient condition \cite{Bose2019}; gas should locally dominate the potential for a non-negligible time period before it gets non-adiabatically expelled, a condition that is more easily satisfied if the density threshold for star formation used in a simulation is high enough \cite{Benitez-Llambay2019, Dutton2019, Dutton2020}, which is physically motivated, because most star formation is observed to occur only in self-gravitating, high-density gas like giant molecular clouds.

Subtleties in the numerical implementation of star formation and feedback, exacerbated by limitations in numerical resolution, historically have prevented a rigorous modeling of core formation. Nearly all baryonic simulations that resolve and model star formation in high-density gas report some degree of core formation in the scale of $0.1 - 1$ kpc in dwarfs \cite{Governato2012,Teyssier2013,DiCintio2014, Onorbe2015, Chan2015, Tollet2016, Read2016, Lazar2020, Jahn2021b}. However, 3 aspects of core formation remain controversial: (1) the link to the star-formation history, (2) the sizes of the cores, and (3) the minimum mass to form a core.

On short timescales ($\lesssim 200$ Myr), the density slope of dark-matter fluctuates between core-like and cusp-like, as gas is expelled and re-cools/re-accretes into the galaxy, which shallows and deepens the overall potential, respectively.
This means that gas-rich star-forming dwarf galaxies should show a diversity of inner-density slopes that correlate with recent star-formation activity \cite{Chan2015, Read2016b, ElBadry2017}.
On longer (cosmological) timescales, the degree of core formation increases with the number of starburst cycles, so dwarf galaxies with more extended star-formation histories should show more prominent cores \cite{Onorbe2015, Chan2015, ElBadry2016, Read2017}; observations indeed suggest this correlation \cite{Read2019}.
Conversely, extended periods without star formation may lead to re-growth of a cusp \cite{Laporte2015}.
However, not all simulations predict such a strong correlation \cite{Benitez-Llambay2019} or the need for a sustained active star formation to show cores \cite{Orkney2021}.

The size of the dark-matter core in some simulations is linked to the half-mass radius of the stars \cite{Chan2015,Read2016,Fitts2017}, while controlled experiments suggest instead that the more concentrated the energy deposition of the feedback is, the more extended the dark-matter core \cite{Benitez-Llambay2019, Burger2021}. With degeneracies in the baryonic modeling of galaxies going hand in hand with structural differences in the stellar component of the simulated galaxies \cite{Sales2010, Scannapieco2012, Agertz2015, Agertz2016}, the predicted sizes of dark-matter cores remains in debate.

Uncertainties also exist on the minimum galaxy mass needed for core formation.
A balance between having enough star formation to affect the potential while still having a relatively low-mass dark-matter halo makes core formation from stellar feedback most efficient at masses comparable to the Large Magellanic Cloud, with $\mstar \sim 10^9$\; \msun\; and halo masses $\sim 10^{11}$\; \msun\; \cite{Governato2012, Penarrubia2012, DiCintio2014,Chan2015,Tollet2016,Lazar2020}. And while for fainter dwarfs this mechanism may lead to smaller and less shallow cores, some analytical arguments imply no core formation in ultra-faint dwarfs \cite{Penarrubia2012}, which agrees with many cosmological simulations that show a `threshold' halo mass for core formation \cite{Madau2014, Onorbe2015, Fitts2017}. On the other hand, different simulation codes recently suggest that ultra-faints should also harbor depleted dark-matter densities \cite{Read2016a, Orkney2021} as a combined result of feedback followed by minor-mergers heating up the dark-matter component and an increased numerical resolution compared to previous simulations. Cores forming all the way down to the ultra-faint regime also seems supported by analytical arguments \cite{Maxwell2015}, highlighting that the minimum mass for core formation from baryonic feedback remains open to debate and perhaps affected by numerical resolution effects.

With firm evidence from several independent numerical codes and analytical models showing that it is possible to form cores at the centers of the dark-matter halos of dwarf galaxies from feedback effects, the core-cusp tension with $\Lambda$CDM is, at this point, only uncertain (as listed in fig.~\ref{fig:fig1}) and awaiting larger samples of observed dwarfs with better observations of inner kinematics. On the theoretical side, a better understanding of the predicted core sizes, correlations with other dwarf properties, and the existence or not of a threshold mass for core formation is also necessary.

However, a closer look into this core-cusp challenge using a compilation of available rotation curves of dwarf galaxies revealed a new, but related, and more challenging tension: observed dwarfs of similar masses ($M_* \geq 10^7\; \rm M_\odot$) show a large \textit{diversity} in the inner shapes of their inferred dark-matter profiles: some are cored, some are consistent with NFW and some are even more concentrated than NFW profiles \cite{Oman2015, Relatores2019} (see Fig.\ref{fig:diversity} for illustration). Moreover, a similar diversity in the dark-matter density of MW satellites has also been found \cite{Zavala2019}, with galaxies like Draco consistent with a steep dark-matter cusp \cite{Read2018,Massari2020} that contrasts the large dark-matter core inferred for, for example, Fornax. 

As discussed above, recent simulations have suggested that baryon-induced core formation is possible and common in dwarfs with medium-to high masses.
However, reproducing this \textit{diversity} of rotation curves, mass ranges and, in particular, including their predicted correlations with other galaxy properties, remains troublesome to all current models \cite{SantosSantos2020a} and therefore a strong tension point between theoretical predictions of $\Lambda$CDM and observations.

Non-circular and out-of-equilibrium motions in observed rotation curves could cause, in principle, an inferred level of diversity similar to observations \cite{Read2016b,Oman2019}. However, the necessarily perturbations to the velocity fields appear inconsistent with the well-behaved rotation curves measured. Overall, we must continue to proceed with caution and apply apples-to-apples comparisons of theory against observations, generating synthetic observations of simulations; for example, in dispersion-dominated galaxies, cusps can be disguised as cores in observations \cite{Genina2018}.
 
Finally, in the most extreme cases of diversity, some observed dwarf galaxies appear to be in fact baryon-dominated or dark-matter poor, for example, DDO 50 and NGC 1613, in which the deficit of dark matter extends well beyond the radius of the stars in discrepancy with baryon-induced cores \cite{Oman2016}. Additionally,  examples of baryon-dominated inner regions have been reported for dispersion-dominated dwarfs such as NGC 6822 \cite{Kirby2014}, ultra-diffuse galaxies DF2 \cite{vanDokkum2018b,Danieli2019} and DF4 \cite{vanDokkum2019} and Antlia II in our own Galaxy \cite{Torrealba2019}.

Barring significant systematics in the observations, the diversity of rotation curves (and enclosed dark-matter mass) is arguably one of the strongest current tensions with theoretical models without a clear and consistent baryonic solution to date \cite{SantosSantos2018,SantosSantos2020a}. In particular, it seems that the same baryonic feedback solutions that appear to solve some of the other tensions that we discuss in this review, also tend to lower the inner dark-matter density in dwarfs too uniformly, driving a strong tension with the degree of diversity observed. This warrants its placement as ``strong tension'' in Fig.~\ref{fig:fig1}.


\section*{Baryonic distribution within dwarf galaxies} 
\label{sec:BARdistribution}

We next discuss the baryonic components of dwarf galaxies, in particular, stellar morphology, identifying an emerging tension: the simultaneous formation of both diffuse and compact dwarfs in simulations.
This is another manifestation of `diversity' in dwarf galaxies.

Most cosmological baryonic simulations of low-mass galaxies that couple star formation to high-density gas predict rapidly varying (`bursty') star-formation \cite{Stinson2007, Gonzalez-Samaniego2014, DiCintio2014, Sparre2017, Bose2019}, although the predicted level of burstiness differs across different simulations \cite{Iyer2020}.
Importantly, because both stars and CDM behave as (effectively) collisionless fluids, stars necessarily experience  similar effects from the fluctuations of the gravitational potential induced by feedback as dark matter does as we described above, with a `breathing mode' of galaxy size fluctuations on short timescales and dynamical heating / puffing out on longer timescales \cite{Read2005, Teyssier2013, ElBadry2016}.
Thus, the phenomenology for stars mirrors that of dark matter, as discussed above.

\begin{figure}[ht!]
\centering
\hspace{-1cm}
\includegraphics[width=0.7\linewidth]{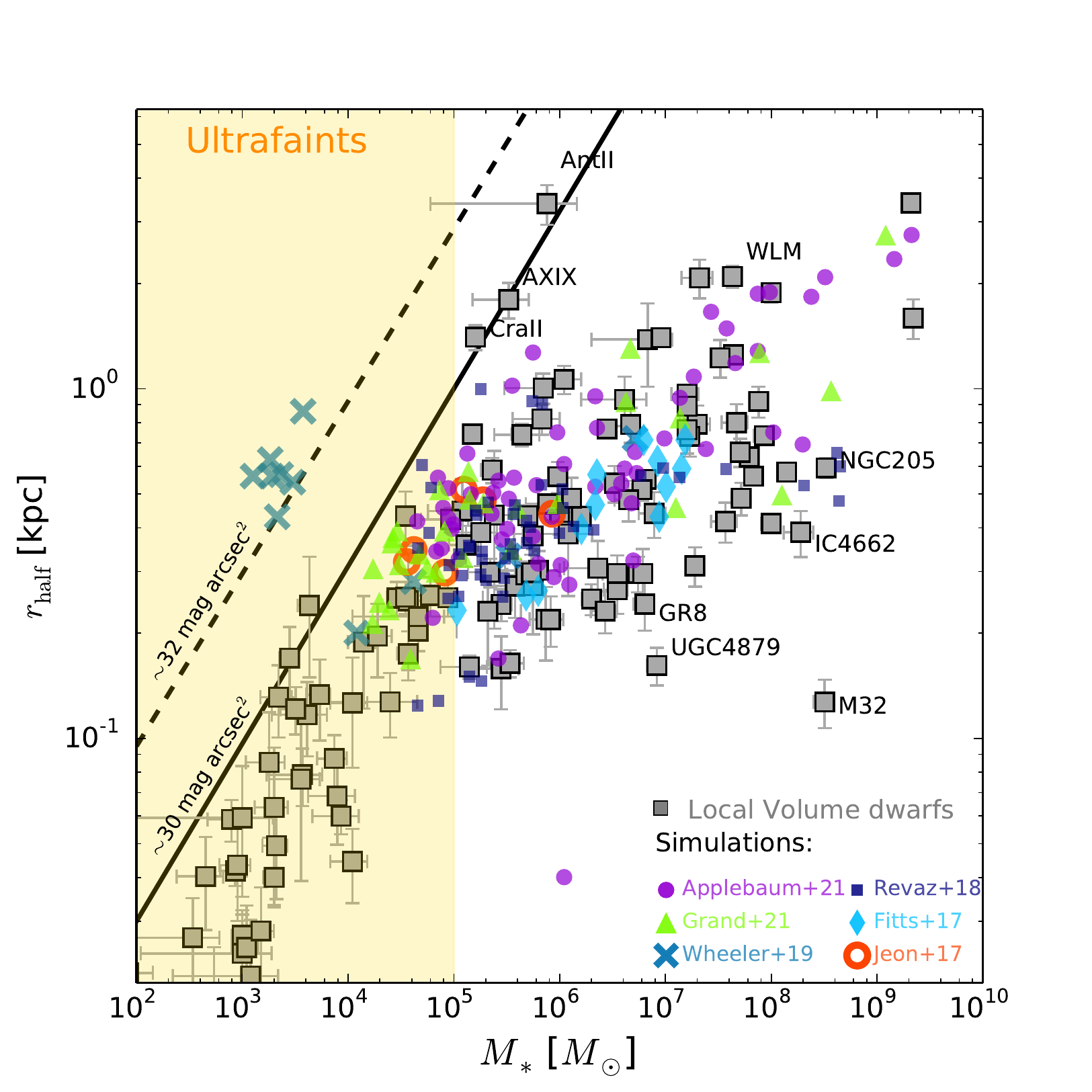}
\caption{
{\bf Dwarf galaxies show a wide range of sizes at fixed stellar mass.}
Current high-resolution simulations reproduce nicely the general trend but, individually, each code shows too little scatter, leaving the diffuse and compact dwarfs underrepresented in the models. The yellow shaded region indicates the ultrafaint dwarf regime. Gray squares show observational data  in the Local Volume dwarfs from ref. \cite{McConnachie2012} updated catalog (satellites and field), Antlia-II data from ref. \cite{Ji2021}. The names of some of the most extreme dwarfs are highlighted. Simulated data is shown in colored symbols: DC Justice (purple circles, ref. \cite{Applebaum2021}), Auriga L2 (green triangles, \cite{Grand2021}), FIRE (sky-blue diamonds, ref. \cite{Fitts2017}), Gadget-2 (open red circles, \cite{Jeon2017}), FIRE low mass (blue crosses, ref. \cite{Wheeler2019}) and GEAR (teal squares, ref. \cite{Revaz2018}). A minimum of $20$ stellar particles apply to Auriga L2 and GEAR simulations for which particle information was made available to us by the authors. The solid black line indicates $\sim 30$ mag arcsec$^2$, approximately the surface-brightness limit in current ultrafaints surveys,  which would leave objects as diffuse as those predicted by, the FIRE-2 simulations ($\sim 32$ mag arcsec$^2$, dashed line) undetected \cite{Wheeler2019}. 
}
\label{fig:sizes}
\end{figure}

As a result of this dynamical heating process for the stars, simulations predict galaxies at $M_* \lesssim 10^9$ \msun\; to be mostly dispersion dominated \cite{Wheeler2017, ElBadry2017}, which at least qualitatively agrees with observations. However, observed dwarfs display a wide range of sizes at a fixed stellar mass, as  Fig.~\ref{fig:sizes} shows for dwarf galaxies in the Local Volume from \cite{McConnachie2012} in gray (data taken from updated list maintained by the author).
 We assume a mass-to-light ratio equal to one to compute stellar mass and calculate the circularized half-light ratio from ref. \cite{McConnachie2012} by multiplying the size along the major axis by $\rm sqrt(1-e)$, with $e$ the ellipticity of the system. Finally, we multiply the circularized projected half-light radius by a constant factor (4/3) to estimate the 3D half mass radius plotted.
The Local Group data is compared with several zoom-in simulations of MW-like halos and their surrounding volume \cite{Applebaum2021, Grand2021} and zoom-in simulations of individual dwarfs \cite{Fitts2017,Jeon2017,Revaz2018,Wheeler2019}. These simulations model the average dwarf population reasonably well, but the intrisic dispersion within each simulation set is appreciably smaller than observations. In particular, diffuse dwarfs like Crater-II, Antlia-II, Andromeda-XIX as well as compact dwarfs such as the dwarf elliptical M32, UGC 4879 and GR8 are underrepresented. 

The problem to form simultaneously diffuse and compact dwarfs may potentially worsen in simulations of higher-density environments like groups and clusters, where diffuse, compact and ultra-compact dwarfs appear in larger numbers \cite{Brodie2011,vanderBurg2016,vanderBurg2017}. Even within the Local Group, while simulations by ref. \cite{Applebaum2021} report no significant issues to match the most extended dwarfs, several other codes (as shown by Fig.~\ref{fig:sizes}) have difficulties matching the most extended objects. In fact, dwarfs as extreme as Andromeda-XIX or Antlia-II are missing in all current simulations. While artificial numerical disruption of such low density systems may be a factor of concern, the systematic lack of diffuse objects in the simulations shown in this figure highlights the need for a better understanding of the physics that set the sizes of the most extended dwarf galaxies.

Some of the difficulties in simulating compact dwarfs may be naturally alleviated by reaching higher numerical resolution \cite{Applebaum2021, Grand2021}, such that numerical softening is at least an order of magnitude smaller than the galaxy itself, so the orbits of stars are followed with more fidelity.
However, even some of the highest-resolution cosmological simulations, such as those in ref. \cite{Wheeler2019}, do not necessarily lead to smaller sizes.
The problem is beyond the artificial softening of gravitational forces in these scales: with burstiness and its associated size fluctuations as inescapable prediction, it is difficult to envision how any compact stellar object can survive without dynamical heating and expansion in current baryonic treatments. Interestingly ref. \cite{Applebaum2021} traces the case of at least one compact dwarf formed with $M_\star \sim 10^7$ \msun\; and half-light radii of only $40$ pc to a heavily tidally stripped subhalo. However such mechanism would not explain some of the compact objects in the Local Volume like UGC 4879 and GR 8, which are in isolation from the MW and M31.

As with core formation, the predicted relation between stellar size/kinematics and star-formation history is observationally testable.
Simulated dwarfs form stars at the highest rate during the gas-contraction phase, when their stellar sizes are small and velocity dispersions are high, while they expand their size in gas-blow out phase when stellar sizes are large and velocity dispersions are low \cite{ElBadry2016, ElBadry2017}.
While existing observations do not support this correlation between stellar size and recent star-formation history \cite{Patel2018}, other observations do support the predicted relationship between kinematics and star-formation history \cite{Hirtenstein2019, Pelliccia2020}.

One possible solution is to consider that burstiness might be over-predicted in current simulations. Attempts at comparing the star-formation timescales predicted to observations indicate that to first order they are consistent, for example, with predictions from the FIRE model \cite{Hopkins2014, Hopkins2018}. However, some works indicate that simulated star-formation histories might be too bursty at $M_\star \leq 10^{7.5}$ \msun\; \cite{Sparre2017, Emami2019, Cignoni2019}. Thus, while in the details the intensity and frequency of star formation in dwarfs is not yet well constrained by the models, the associated breathing mode seems fundamental to establishing observed negative metallicity gradients \cite{ElBadry2016, Mercado2021} (at least in some models like FIRE), dark-matter cores and even stellar halos in dwarfs \cite{Stinson2009}.

Understanding how to form compact stellar systems while simultaneously preserving the adequate level of burstiness to reproduce the observed properties of the more extended and less dense dwarfs remains a key challenge to galaxy formation models within \lcdm. We list the diversity of luminous sizes of dwarf galaxies as a weak tension in Fig.~\ref{fig:fig1} and highlight that photometric/kinematical studies of individual stars in dwarfs as well as integrated fluxes as proposed in ref. \cite{Emami2019} might hold the key to observationally constrain how bursty star-formation histories are in dwarf galaxies.

\section*{The too big to fail problem} 
\label{sec:TBTF}

As highlighted by Refs. \cite{Boylan-Kolchin2011, Boylan-Kolchin2012}, the dark-matter mass --inferred indirectly from the stellar kinematics of stars within the half-light radius-- for the most massive observed satellites of the MW is typically smaller than those of the massive subhaloes (which should then host these galaxies) of the simulated MW halos in the Aquarius dark-matter-only simulations \cite{Springel2008}. One solution is to require that several massive subhalos (V$_{\rm peak} \gtrsim 30\,\kms$) in simulations must be completely dark, but this is problematic, because such subhalos are massive enough that their gas should have cooled and formed stars; in other words, they are `too big to fail' at hosting galaxies. Spectroscopic measurement of stellar velocity dispersions of dwarf in ref. \cite{Tollerud2012} and \cite{Collins2013} argued for a similar too big to fail (TBTF) problem in dwarf spheroidal satellites of M31, noting that the more compact dwarf ellipticals do not suffer from this problem.

Although originally stated as a tension for satellites, the TBTF problem also was found in central (field) galaxies within the Local Group \cite{Garrison-Kimmel2014} and later generalized to other isolated dwarf galaxies in the nearby Universe where the analysis of their rotation curves indicated halo masses that are lower than predicted from abundance matching relations \cite{Ferrero2012, Papastergis2016}. This solidified TBTF as a tension in the field environment. Since the original discussion of the TBTF problem, several solutions have been proposed based on the study of different cosmological simulations. We outline below the key proposed mechanisms to address the TBTF problem, some of which pertain only to the `satellite' version of the problem.

First of all, the TBTF problem for satellites could be naturally alleviated, before invoking any baryonic effect, by lowering the mass assumed for the MW-mass host halo, given the predicted dependence of subhalo numbers on this in $\Lambda$CDM \cite{WangFrenk2012, Vera-Ciro2013}. Although still within observational constraints, this solution then suggests that the true mass of the MW halo lays in the lower-half of the currently allowed estimates, which may conflict with the presence of a massive satellite such as the Large Magellanic Cloud or the large velocity of Leo-I \cite{Boylan-Kolchin2013}. Halo-to-halo scatter on the subhalo content is also an important factor to consider \cite{JiangVDB2015,Fielder2019}. For example, as shown in ref. \cite{WangFrenk2012}, the Aquarius halos used to first pose the TBTF problem have all above average number of subhalos. The extension of this argument also applies to the TBTF in the field in the Local Group; such that the number of halos above a given mass threshold depends on the total mass of the Local Group including mass outside MW and M31 virial radii \cite{Fattahi2020}.

Considering baryons introduces several other solutions. First, as discussed earlier, most high-resolution baryonic simulations predict the formation of dark-matter cores, which alleviates the TBTF problem by reducing the dark-matter mass in the inner region without requiring dwarf galaxies (satellites or field) to reside in lower-mass halos. This mechanism has been highlighted to contribute to the solution of the TBTF problem in the middle- to high-mass range of classical dwarfs, where core formation from baryonic processes is most effective \cite{Brooks2014,Brook2015,Wetzel2016,Papastergis2016,Garrison-Kimmel2019}. Additionally, modeling the baryons in MW-like simulations revealed an important factor to resolve the TBTF problem in satellites: the gravitational potential from the central galaxy causes enhanced tidal stripping in satellites that is not present in dark-matter-only simulations, making subhalos more susceptible to mass loss and enhancing disruption of dwarf galaxies \cite{Donghia2010, Penarrubia2010, Brooks2014, Wetzel2016, Garrison-Kimmel2017, Errani2017, Kelley2019, Samuel2020}. This mechanism contributes to addressing the TBTF problem for satellites (but not in the field) at all masses, thus it is particularly important for low-mass dwarf galaxies, where core formation is less efficient.

A more subtle factor to consider is that the total halo mass (or similarly $\vmax$) of halos (and subhalos) in baryonic simulations are lower than their matched counterparts in dark-matter-only simulations.
This is generally true regardless of whether the baryonic simulations produce dark-matter cores or not \cite{Sawala2013,Brooks2014}, for two reasons.
First, the (external) UV background and (internal) stellar feedback remove a significant fraction of the baryons from $V_{\rm max} < 50$ km/s dwarf halos. Second, this lower mass throughout most of the cosmic time results in reduced cosmic accretion. This relatively small reduction of halo mass has a considerable effect on reducing the severity of TBTF for field and satellite galaxies, given the steep shape of the (sub)halo mass function \cite{Brooks2014,Sawala2016,Fattahi2020}.

In summary, there is a consensus among current cosmological simulations of MW/M31-mass halos that \textit{there is no TBTF problem for MW and M31 satellites}, regardless of whether the simulations produce cuspy or cored dark-matter profiles. We therefore report no apparent tension between observations and predictions in the context of TBTF {\it for satellites} in the Local Group.

However, the situation is less clear for the TBTF problem in the field.
Several works have argued that alongside the baryonic effects discussed above, including an adequate comparison between simulations and observations that takes into account observational biases and techniques, are able to reconcile the predicted and observed velocity function of isolated gas-rich dwarfs as given by HI-width data \cite{Maccio2016, Brooks2017, Verbeke2017, Chauhan2019}. This solution to the TBTF problem in the field relies partially on the level of turbulence in the ISM of dwarf galaxies and also on the formation of dark-matter cores, the details of which, as discussed above, are not fully settled. Moreover, with large uncertainties in the incompleteness of data and total mass of the LG, it is not clear whether massive `unaccounted-for' halos in the LG field is a source of tension, and whether or not the predicted small velocity dwarfs will be accounted for in upcoming HI observational surveys. There still remain several observed dwarf galaxies with full rotation curve data, such as DDO 50 and IC 1613 among others \cite{Ferrero2012, Oman2016}, suggesting a dark-matter halo substantially less massive than predicted by abundance matching models, along the direction of the original TBTF claims. More recently, several ultra-diffuse dwarf galaxies in the field have also been found to have lower dark-matter masses than theoretically expected \cite{ManceraPina2019,ManceraPina2021}. We therefore assess this problem as a weak tension in Fig.~\ref{fig:fig1}. Investigations into the diversity of rotation curves (or core-cusp problem), as well as future discovery of nearby field dwarfs using up-coming surveys, such as Rubin Observatory, will be fundamental to assess the level of tension, if any, for TBTF in the field.
\section*{Satellite dwarf galaxies: in the Local Group and around nearby Milky Way analogs}
\label{sec:sats}

We finally review three additional tests of simulation predictions for dwarf galaxies that are satellites around a MW-mass galaxy (within $\approx 300 - 400$ kpc).

A long-standing challenge for cosmological simulations has been achieving sufficient resolution to model the spatial distribution of satellite galaxies within a MW-mass halo, without suffering from artificial numerical `over-merging' \cite{Moore1996, Klypin1999, WetzelWhite2010, vandenBosch2018b}.
Cosmological zoom-in simulations that model only dark matter achieved high numerical resolution \cite{Diemand2008, Springel2008, Stadel2009}, but their lack of baryons and a MW-mass galaxy limited their accuracy \cite{TaylorBabul2001, Donghia2010, GarrisonKimmel2017, Sawala2017}.
Cosmological zoom-in simulations that include baryons now achieve sufficient resolution to match the radial distribution of satellite dwarf galaxies (at least at $\mstar \gtrsim 10^5$ \msun) as observed around the MW, M31, and nearby MW-mass analogs \cite{Kelley2019, Samuel2020, Libeskind2020, Grand2021, Font2021}.
Thus, while a detailed understanding of physical versus numerical effects remains ongoing and essential \cite{vandenBosch2018b}, in our evaluation current simulations of MW-mass galaxies show reasonable agreement with observed radial distance distributions of satellite dwarf galaxies (however, see ref. \cite{Carlsten2020}).

More significant tension has persisted between simulations and observations regarding the 3-D spatial and 3-D velocity distribution of satellites.
Nearly all of the satellites around the MW \cite{LyndenBell1976, Kroupa2005, Pawlowski2012, Fritz2018, PawlowskiKroupa2020} and about half of the satellites around M31 \cite{Conn2013, Ibata2013} are in a kinematically coherent, thin planar distribution.
Some nearby galaxies show planar distributions of satellites as well, such as Centaurus A \cite{Muller2018, Muller2021}, M101 \cite{Muller2017} and the MATLAS sample of massive elliptical galaxies \cite{Heesters2021}. Many works have argued that the relative thinness of these satellite planes, and their kinematic coherence, strongly disagree with predictions from cosmological simulations, though with considerable debate 
\cite{Metz2008, Cautun2015,  Buck2016, Pawlowski2018}.

\begin{figure}[ht]
\centering
\hspace{-1 cm}
\includegraphics[width = 0.86 \linewidth]{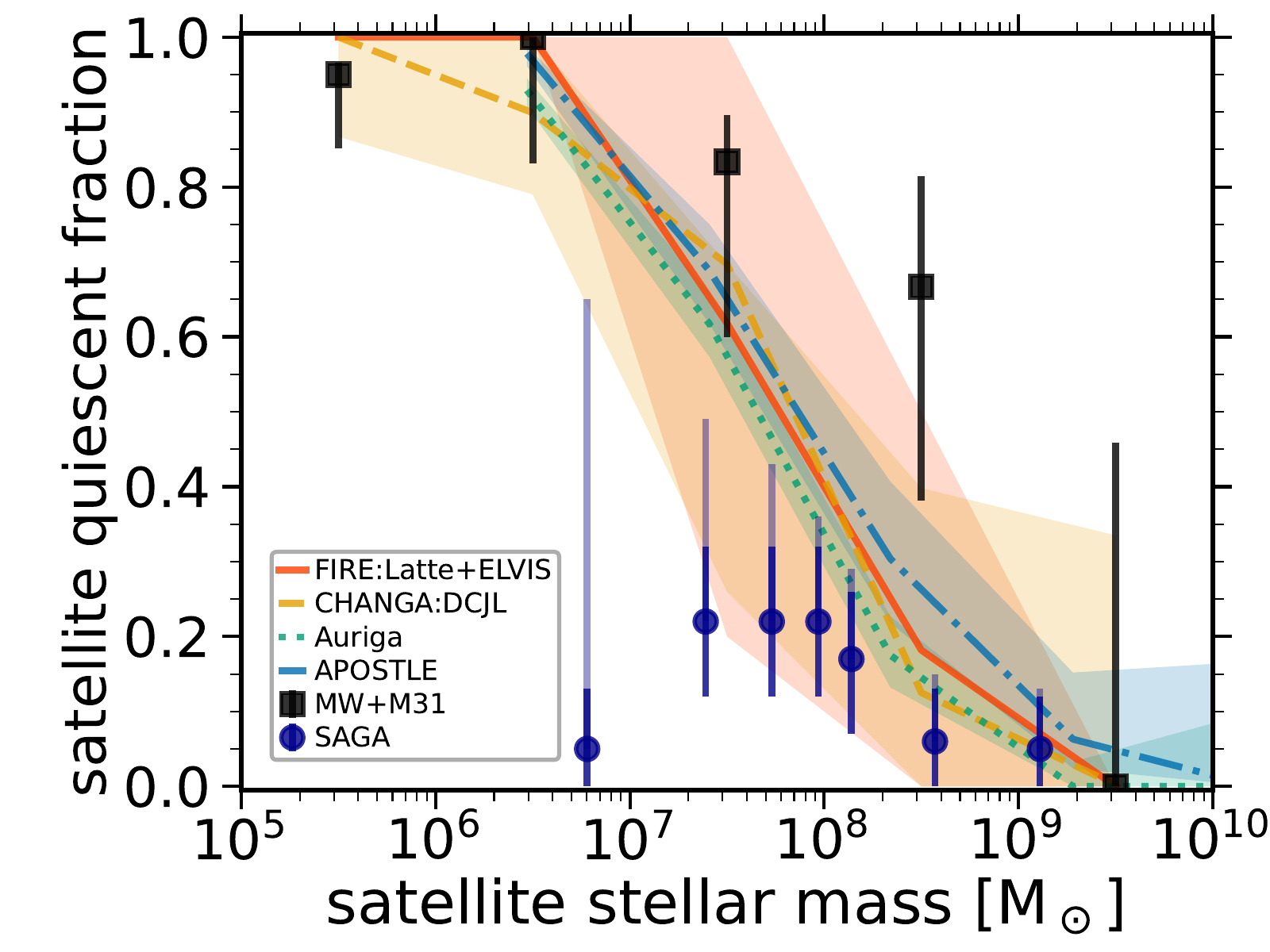}
\caption{
\textbf{Fraction of satellite dwarf galaxies that are quiescent (non-star-forming) versus stellar mass, in observations and simulations of Milky Way-mass galaxies,}
adapted from ref. \cite{Samuel2022}.
Black points show observed satellites around the MW and M31, adapted from ref. \cite{Wetzel2015b} using the observational compilation in ref. \cite{Putman2021}.
Blue points show observed satellites around 36 nearby MW-mass galaxies from SAGA \cite{Mao2021}.
In both cases, error bars show 68\% uncertainty from observed counts, while for SAGA, light bars additionally show the maximal spectroscopic incompleteness correction in their survey.
Lines show simulations of MW-mass galaxies: FIRE-2 Latte + ELVIS suites (red; ref. \cite{Jahn2021, Samuel2022}, shading indicates host-to-host scatter), CHANGA DC Justice League suite (orange; ref. \cite{Akins2021}, shading indicate scatter across hosts and satellite counts), Auriga (green) and APOSTLE (purple) suites (ref. \cite{Simpson2018, Karunakaran2021}, shading shows scatter from counts alone).
At $\mstar \gtrsim 10^9$ \msun, both observed and simulated quiescent fractions broadly agree near 0.
Down to $\mstar \sim 10^7$ \msun, all simulations lie between the MW+M31 and SAGA, though agree better with the former.
At $\mstar \lesssim 10^7$ \msun, all simulations predict quiescent fractions near unity, which agrees well with MW+M31 but disagrees with SAGA.
}
\label{fig:qufrac}
\end{figure}

The nature of these planes of satellites has persisted as one of the strongest tensions between theory and observations.
Refs. \cite{Pawlowski2021} and \cite{BoylanKolchin2021} provide  extensive recent commentary on this topic; here we mention only two aspects recently explored that likely play an important role in comparing simulation predictions against observations of the MW and M31.
First, simulations show that the presence of a massive satellite like the LMC (or M33/M32) significantly can boost the planarity of the satellite population \cite{Samuel2021}, by accreting many satellite together on a similar orbit \cite{LiHelmi2008, Donghia2008, Sales2011, Deason2015, SantosSantos2021} and focusing the planarity of existing satellites \cite{GaravitoCamargo2021}.
Second, the planar structures of dwarf galaxies around the MW, M31, and the Local Group as a whole show some degree of alignment \cite{PawlowskiMcGaugh2014}, which suggests importance in modeling the larger-scale cosmological structure around the Local Group \cite{Neuzil2020, Libeskind2020}.

A compelling emerging tension for satellite dwarf galaxies regards their star formation and gas content.
Theory predicts that most dwarf galaxies with $\mstar \gtrsim 10^{5-6}$ \msun retain their cold gas after cosmic reionization and thus remain star-forming \cite{Fitts2017}, \textit{if} they do not become a satellite in a larger (MW-mass) host halo.
Indeed, nearly all observed isolated (non-satellite) dwarf galaxies are star-forming \cite{Geha2012}, with only 3 known exceptions \cite{Makarov2012, Karachentsev2015, Polzin2021}.
Furthermore, nearly all dwarf galaxies in the Local Group beyond the halo radius ($\gtrsim 300$ kpc) of the MW and M31 are star-forming, but by contrast, nearly all satellites of the MW and M31 are quiescent, with no gas and no star formation \cite{McConnachie2012, Weisz2015, Wetzel2015b, Putman2021}.

This stark contrast for satellite versus central dwarf galaxies in the Local Group suggests that the environmental effects of a MW-mass halo are efficient in stripping gas (likely via ram pressure) out of satellites and quenching their star formation.
Indeed, as Figure~\ref{fig:qufrac} shows, current cosmological zoom-in simulations of MW-mass galaxies generally show efficient gas stripping and thus high quiescent fractions for satellites at $\mstar \lesssim 10^8$ \msun, which are broadly consistent with the MW and M31 \cite{Simpson2018, GarrisonKimmel2019b, Akins2021, Joshi2021, Karunakaran2021, Jahn2021, Font2021b, Samuel2022}; though see ref. \cite{Hausammann2019} for a differing perspective.

However, recent observations of satellites beyond the Local Group suggest a strikingly different picture.
The SAGA survey \cite{Geha2017, Mao2021} has published quiescent fractions for 127 satellites at $\mstar \gtrsim 10^7$ \msun around 36 nearby MW analogs --- much more cosmologically representative than just the MW and M31 of the Local Group.
As Figure~\ref{fig:qufrac} shows, SAGA finds that nearly all satellites are star-forming, with only $\lesssim 20\%$ quiescent at all masses they probe, significantly lower (even considering potential incompleteness effects) than the $\gtrsim 70\%$ quiescent fractions at these masses around the MW and M31.
At face value, these SAGA results upend the long-standing expectation that MW-mass halos are efficient at stripping gas and quenching star formation in satellite dwarf galaxies.

As Figure~\ref{fig:qufrac} also shows, the quiescent fractions of satellites in SAGA are substantially lower than all current cosmological zoom-in simulations at $\mstar \lesssim 10^8$ \msun.
One possibility is significant incompleteness of (diffuse) quiescent galaxies in the SAGA survey, as ref. \cite{Font2021b} suggest; although, if true, this would seem to require the existence of quiescent dwarf galaxies at lower surface brightness than observed in the Local Group.
Taken at face value, the SAGA results imply a new tension: that simulations of MW-mass halos are in fact \textit{too efficient} at stripping star-forming gas out of satellite dwarf galaxies (as suggested by the simulations results of ref. \cite{Hausammann2019}).
Thus, these SAGA results raise new questions:
Why have the MW and M31 been so efficient in quenching star formation in their satellites?
Is the Local Group a cosmological outlier in this sense?
Do cosmological simulations over-predict the efficiency of gas stripping and star-formation quenching for satellites in a typical MW-mass halo?

In summary, simulations show reasonable agreement with the radial distance distributions of satellites, but as we list in Fig.~\ref{fig:fig1}, significant tension persists regarding the planarity of the 3-D distribution, and the quenching of star formation in satellites presents a new tension, though more work is needed to understand the uniqueness of the Local Group and the completeness of surveys like SAGA.

\section*{Future challenges}

Three factors will drive progress in the near future for theoretical studies of dwarf galaxies: (1) improvements in the numerical power of simulations, propelled by optimized codes and higher-performance computing clusters; (2) implementations of additional physics and improved implementations of processes already modeled in the interstellar medium of dwarf galaxies; (3) new observational constraints on the population and star-formation histories on small timescales of dwarf galaxies in both the early Universe and ultra-faint galaxies at the present-day.
This includes the detection and characterization of the population of completely dark (sub)halos (without stars or gas), which is one of the strongest untested predictions of galaxy formation in $\Lambda$CDM plus baryons.

Improvements on numerical resolution importantly will enable the exploration of more diverse cosmic environments, including those of groups and galaxy clusters, where dwarf galaxies display more extreme ranges of star formation histories and morphologies, including both a numerous population of ultra-diffuse and ultra-compact dwarfs. Mighty efforts are already underway \cite{Tremmel2019, Pillepich2019, Nelson2019, Dubois2020}, but more resolution is desirable to resolve fainter dwarfs along with their sizes and inner baryonic plus dark matter structure.

Frontier simulations will include a richer set of physical processes. For example, feedback from black holes has been confirmed observationally in several dwarf galaxies with masses $M_*\sim 10^8-10^9\; \rm M_\odot$ \cite{Penny2018,Dickey2019,Manzano-King2019,Greene2020}, while most simulations of dwarf galaxies do not include the physics of black holes (although some efforts are underway, see refs. \cite{Bellovary2019,Sharma2020,Koudmani2021,Volonteri2020}). Magnetic fields and their interaction with cosmic rays likely affects the ability of dwarf galaxies to form stars and drive outflows \cite{Uhlig2012,Pakmor2016,ChenBryan2016,Bustard2020,Dashyan2020,Hopkins2021,Semenov2021}, but these processes only now are starting to be modeled in dwarf galaxies, with significant numerical development to come. As telescopes peek deeper into the early Universe, improved treatments for reionization and the evolution of the UV background, effects of radiation via radiative transfer, low-metallicity star formation and the first generation (Pop III) stars will become key to making robust theoretical predictions, especially for ultra-faint dwarf galaxies \cite{Wise2012, Johnson2013, Paardekooper2013, Wise2014, Dayal2018, Jeon2017}. Alongside the improvement of the physics, future studies should also address the effects of randomness and chaotic behavior on solving the differential equations at the heart of simulations on the scale of dwarf galaxies \cite{Genel2019, Keller2019}.

Observationally, beyond volume-complete census for fainter dwarfs being in the horizon with upcoming telescopes like Rubin Observatory, ELT or the Roman Space Telescope, measuring the satellite mass functions around low-mass primaries in the field may represent an attractive and more efficient alternative to reach the regime of ultra-faint dwarfs where most theoretical predictions differ. In fact, because dwarf galaxies are also expected to host their own population of satellites \cite{Sales2013,Wheeler2015,Dooley2017,Jahn2019}, and they are more abundant cosmologically than MW-mass galaxies, they might represent ideal candidates to survey their satellite content and put strong constraints on the abundance and properties of ultra-faint dwarfs. Several promising observational efforts on this direction might be able to add exciting  constraints in the near future \cite{Drlica-Wagner2021,Carlin2021,Davis2021,Roberts2021,Muller2020} which should inform current baryonic galaxy formation models \cite{Munshi2019}.

Dwarf galaxies stand strong as powerful cosmological probes.  Contrasting their observed properties to baryonic simulations will continue to improve our galaxy formation models and their numerical implementations. But dwarfs are also key to understanding the nature of dark matter: if the current tensions highlighted in this article --and any to be discovered in the future--remain unanswered by improved baryonic treatments coupled to a CDM scenario, this will advocate strongly for the need for an alternative dark-matter model beyond $\Lambda$CDM.
Said differently, understanding and accurately modeling baryonic effects is a necessary prerequisite to any rigorous test of dark matter in the regime of dwarf galaxies.

\section*{Acknowledgements} 

We would like to thank the Yves Revaz and the two anonymous referees for their constructive comments that helped improved the first version of this review. 
We thank Jenna Samuel, Ananthan Karunakaran, Yvez Revaz, Rob Grand and Ferah Munshi for sharing simulation data.
We also thank Kyle Oman for generating Fig. 3 of this review.
LVS is grateful for financial support from NASA ATP 80NSSC20K0566, NSF AST 1817233, 2107993 and NSF CAREER 1945310 grants. 
AW received support from: NSF grants CAREER 2045928 and 2107772; NASA ATP grants 80NSSC18K1097 and 80NSSC20K0513; HST grants AR-15057, AR-15809, GO-15902, GO-16273 from STScI; a Scialog Award from the Heising-Simons Foundation; and a Hellman Fellowship.
AF is supported by a UKRI Future Leaders Fellowship (grant no MR/T042362/1).

\section*{Data availability}

This review has no new data to report.

\section*{Code availability}

This article presents a review of presented discussions and results in the literature and has not required the development of any new codes.

\section*{Author contributions statement}

All the authors in this review have made substantial contribution to the discussion, writing and editing of all sections in the text. LVS is responsible for Fig. 1 and 4, AW for Fig. 5, and AF for Figs. 2.

\section*{Competing interests}

The authors declare no competing interests.

\bibliography{main}

\end{document}